\documentstyle{cupconf}
\newcommand{\apgt}{\ {\raise-.5ex\hbox{$\buildrel>\over\sim$}}\ }
\newcommand{\aplt}{\ {\raise-.5ex\hbox{$\buildrel<\over\sim$}}\ }

\begin{document}
\title[Type Ia Supernovae and their implications for cosmology]
{Type Ia Supernovae and their implications for cosmology}
\author[M.\ Livio]{M\ls A\ls R\ls I\ls O\ns L\ls I\ls V\ls I\ls O}
\affiliation{Space Telescope Science Institute, 3700 San Martin Drive,
Baltimore, MD 21218}
\maketitle

\begin{abstract}
Models for Type~Ia Supernovae (SNe~Ia) are reviewed. It is shown that
there are strong reasons to believe that most SNe~Ia represent thermonuclear
disruptions of C--O white dwarfs, when these white dwarfs reach the
Chandrasekhar limit and ignite carbon at their centers.

Different progenitor scenarios are reviewed critically and the strengths and
weaknesses of each scenario are presented in detail. It is argued that
theoretical considerations currently favor single-degenerate models, in
which the white dwarf accretes from a subgiant or giant companion. However,
it is still possible that more than one progenitor class contributes to the
observed sample. The relation of the different models to the use of SNe~Ia
for the determination of cosmological parameters is discussed. It is shown
that while the observed diversity of SNe~Ia may argue for the existence of
different progenitor classes, this does not affect the interpretation of an
accelerating expansion of the universe.

Crucial observational tests of the conclusions are suggested.
\end{abstract}

\section{Introduction}
During the past three years two groups (Perlmutter et~al.\ 1997; Schmidt
et~al.\ 1998)
have presented strong evidence that the expansion of the universe is
accelerating rather than decelerating
(Riess et~al.\ 1998; Perlmutter et~al.\ 1998, 1999; and see Livio 1999 for a
perspective). This surprising result comes from distance measurements to
more than fifty supernovae Type~Ia in the redshift range $z=0.1$
to $z=1$. The results are consistent with the cosmological constant (or
vacuum
energy) contributing to the total energy density about 60--70\% of the
critical
density, which in turn, is consistent with recent measurements of the
anisotropy of
the cosmic microwave background (e.g.\ Miller et~al.\ 1999; Wilson et~al.\
1999;
Mauskopf et~al.\ 1999).

This unexpected finding, as well as the use of supernovae Type~Ia to
measure the Hubble constant (e.g.\ Sandage et~al.\ 1996; Saha et~al.\
1997), have focused the attention again on the
frustrating fact that in spite of decades of research, the exact nature
of the progenitors of supernovae Type~Ia remains unknown. Until this
problem is solved, one cannot be fully confident that supernovae at
higher redshifts are not somehow different from their low redshift
counterparts. In the present review I therefore examine critically models
for supernovae Type~Ia and their progenitors. Other recent reviews include
Branch et~al.\ (1995), Livio (1996a; 2000), Renzini (1996), Iben (1997), and
see H\"oflich \& Dominguez, these procedings.

\section{SNe Ia characteristics and the basic model}

The {\it defining\/} characteristics of supernovae Type~Ia (SNe~Ia) are
both spectral: (i)~the {\it lack\/} of lines of hydrogen, and (ii)~the {\it
presence\/} of a strong red Si~II absorption feature ($\lambda$6355
shifted to $\sim6100$~\AA).

Once defined as SNe~Ia, the following are several of the important {\it
observational characteristics\/} of the class which may help in the
search for progenitors:
\begin{enumerate}
\item[(1)] {\it Homogeneity\/}: Until very recently, it has generally been
claimed that more than 80\% of all SNe~Ia form a homogeneous class (see
however (2) below) in terms of their {\it spectra\/} (e.g.\ Branch, Fisher,
\& Nugent 1993), {\it light curves\/}, and {\it peak absolute magnitudes\/}.
The latter are given by
\begin{equation}
M_{\rm B}\simeq M_{\rm V}\simeq-19.30(\pm0.03)+5\log (H_0/60~{\rm
km~s}^{-1}~{\rm
Mpc}^{-1})
\end{equation}
with a dispersion of $\sigma(M_{\rm B})\sim\sigma(M_{\rm V})\sim0.2$--0.3
(Hamuy et~al.\ 1996a; Tamman \& Sandage 1995; and see Branch 1998 for a
review).
\item[(2)] {\it Inhomogeneity\/}: Some differences in the spectra and light
curves have been known to exist for a while (e.g.\ Hamuy et~al.\ 1996b). In
terms of explosion
strength, SNe~Ia have traditionally been roughly ordered as follows: SNe~Ia
like SN~1991bg and SN~1992K
represent the weakest events, followed by weak events like 1986G,
followed by about 80\% of all SNe~Ia which are called ``normals'' (or
sometimes
``Branch normals''), to the stronger than normal events like SN~1991T. In a
very recent work, however, Li et~al.\ (2000) find indications for a
considerably higher {\it peculiarity\/} rate, a total of ($39\pm10$)\%; of
which ($19\pm7$)\% and ($21\pm7$)\% are SN~1991bg-like and SN~1991T-like
objects respectively.
\item[(3)] The {\it luminosity function\/} of SNe~Ia was found in earlier
studies to decline very steeply on
the bright side (e.g.\ Vaughan et~al.\ 1995). Since selection effects
cannot prevent the discovery of SNe which are brighter than the
``normals'' (unless they occur preferentially in high extinction regions),
this is usually taken to imply that {\it the normals are essentially the
brightest\/}. The recent study of Li et~al.\ (2000) seems to show, however,
that the luminosity function is relatively flat at both the overluminous and
underluminous ends.
\item[(4)] Near maximum light, the spectra are characterized by {\it high
velocity\/} (8000--30,000 km~s$^{-1}$) {\it intermediate mass
elements\/} (O--Ca). In the late, nebular phase, the spectra are
dominated by forbidden lines of iron (e.g.\ Kirshner et~al.\ 1993;
Wheeler et~al.\ 1995; Ruiz-Lapuente et~al.\ 1995; G\'omez et~al.\ 1996;
Filippenko 1997).
\item[(5)] Fairly young populations appear to be very efficient at producing
SNe~Ia (e.g.\ they tend to be associated with spiral arms in spirals;
Della Valle \& Livio 1994; Bartunov, Tsvetkov \& Filimonova 1994), but
relatively old populations ($\tau\apgt4\times10^9$~yr) can also produce
them. In particular, {\it SNe~Ia do occur in ellipticals\/} (e.g.\
Turatto, Cappellaro \& Benetti 1994). In fact, the rates of SNe~Ia in
ellipticals appear to be similar to those in spirals, $\sim0.18SNu$ (where
$SNu=1SN(100$~yr)$^{-1} \left(10^{10}L^B_{\odot}\right)^{-1}$; Turatto,
Cappellaro \& Petrosian 1999). This immediately implies that
{\it SNe~Ia are not caused by the core collapse of stars more massive
than 8~M$_{\odot}$\/}.
\item[(6)] There exist a number of correlations between different pairs of
observables (see e.g.\  Branch 1998 for a review). Of these, the most
frequently used in the context of determinations of cosmological
parameters is the correlation between the {\it absolute magnitude and the
shape of the light curve\/}. Basically, brighter SNe~Ia decline more
slowly. A parameter commonly used to quantify the light curve shape is
$\Delta m_{15}$ (Phillips 1993), the decline in magnitudes in the
$B$~band during the first 15~days after maximum light. Hamuy et~al.\
(1996a) find slopes $dM_{\rm B}/d\Delta m_{15}=0.78\pm 0.17$, $dM_{\rm
V}/d\Delta m_{15}=0.71\pm 0.14$, and $dM_{\rm I}/d\Delta m_{15}=0.58\pm
0.13$. Using a stretch-factor $s$ (Perlmutter et~al.\ 1997), one can
write $M_{\rm B}=M_{\rm B}(s=1)-\alpha *(s-1)$, with $M_{\rm B}
(s=1)=-19.46$ (e.g.\ Sandage et~al.\ 1996), and $\alpha=1.74$
(Perlmutter et~al.\ 1999). Sophisticated techniques for using the different
correlations in distance determinations have been developed (e.g.\ Riess
et~al.\ 1996, 1998).
\end{enumerate}

The above characteristics can be augmented by the following suggestive
facts:
\begin{enumerate}
\item[(1)] The {\it energy\/} per unit mass, $1/2(\sim10^4$~km~s$^{-1})^2$,
is
of the order of the one obtained from the conversion of carbon and
oxygen to iron.
\item[(2)] The fact that the event is explosive suggests that {\it
degeneracy\/} may play a role.
\item[(3)] The spectrum appears to contain no hydrogen.
\item[(4)] The explosions can occur with long delays, after the cessation of
star formation.
\end{enumerate}

All the properties above have led to one agreed upon model: {\it SNe~Ia
represent thermonuclear disruptions of mass accreting white dwarfs\/}.

It is interesting that there exists a unanimous consensus on this model
in spite of the fact that the essence of flame physics, burning front
propagation, and the details
of the (presumed) transition from deflagration to detonation (in particular
the
density at which the transition occurs), which are at the heart of the
model, remain as major unsolved problems (e.g.\ Khokhlov, Oran \& Wheeler
1997; Woosley 1997; Reinecke, Hillebrandt \& Niemeyer 1998; and see
H\"oflich \& Dominguez, and Khokhlov, these proceedings). In fact, given
these uncertainties, it is
almost difficult to understand how the entire family of SNe~Ia light
curves can be fitted essentially with one parameter (e.g.\ Perlmutter
et~al.\ 1997), although it is possible that all SNe~Ia explode at the
same WD mass (see \S4), and that the entire observed diversity stems
from different $^{56}$Ni masses.

\section{Why is identifying the progenitors important?}

The fact that we do not know yet what are the progenitor systems of some
of the most dramatic explosions in the universe has become a major
embarrassment and one of the key unsolved problems in stellar and binary
star evolution.
There are several important reasons why identifying the progenitors has
become more crucial than ever:
\begin{enumerate}
\item[(i)] The use of SNe~Ia as one of the main ways to determine
key cosmological parameters like $H_0$, and the contributions to the energy
density (by matter and by the cosmological constant) $\Omega_{\rm M}$,
$\Omega_{\Lambda}$ requires an understanding of the evolution of the
luminosity,  and the SN rate with cosmic epoch.
Both of these depend directly on the nature of the progenitors.
\item[(ii)] Galaxy evolution depends on the radiative, kinetic energy,
and nucleosynthetic output of SNe~Ia (e.g.\ Kauffmann, White \&
Guiderdoni 1993).
\item[(iii)] Due to the uncertainties that still exist in the explosion
mechanism itself, a knowledge of the initial conditions and of the
distribution of matter in the environment of the exploding star are
essential for the understanding of the explosion.
\item[(iv)] An unambiguous identification of the progenitors, coupled
with observationally determined SNe~Ia rates can help to place
meaningful constraints on the theory of binary star evolution (e.g.\
Livio 1996b; Li \& van den Heuvel 1997; Yungelson \& Livio 1998; Hachisu,
Kato \& Nomoto 1999). In particular, a semi-empirical determination of
the elusive common-envelope-ejection efficiency parameter, $\alpha_{\rm
CE}$, may
be possible (e.g.\ Iben \& Livio 1993).
\end{enumerate}

\section{Refinements to the basic model}

The basic model for SNe~Ia (that essentially all researchers in the field
agree upon) is that of a thermonuclear disruption of an accreting white
dwarf (WD). However, additional refinements to the model are possible on
the basis of existing observational data and theoretical models. These
refinements still do not involve the question of the {\it progenitor
systems\/}. Rather, they address the question of the WD {\it
composition\/}, and of its {\it mass\/} at the instant of explosion.

\subsection{The composition of the exploding WD}

In principle, the WD that accretes to the point of explosion could be
composed of He, of C--O, or of O--Ne. Let us examine these possibilities
one by one.
\begin{enumerate}
\item[(i)] {\it He WDs\/}: Helium WDs have typical masses that are smaller
than $\sim0.45$~M$_{\odot}$ (e.g.\ Iben \& Tutukov 1985). While if
accreting,
these He WDs can explode following central He ignition at
$\sim0.7$~M$_{\odot}$, the composition of the ejected matter in this
case will be that of He, $^{56}$Ni and decay products (e.g.\ Nomoto \&
Sugimoto
1977; Woosley, Taam \& Weaver 1986). This is entirely inconsistent with
observations (observational characteristic~(4) in \S2). Therefore, {\it
He WDs certainly do not produce the bulk of SNe~Ia\/}.
\item[(ii)] {\it O--Ne WDs\/}: Oxygen--Neon WDs form in binaries from
main sequence stars of $\sim10$~M$_{\odot}$, although the precise range
which allows formation is somewhat uncertain (e.g.\ Iben \& Tutukov
1985; Canal, Isern \& Labay 1990; Dominguez, Tornamb\'e \& Isern
1993). These systems are probably not numerous enough to constitute the
main channel of SNe~Ia (e.g.\ Livio \& Truran 1992; Livio 1993). It is also
generally expected that O--Ne WDs that manage to accrete enough material
to reach the Chandrasekhar limit will produce (via electron capture)
preferentially
accretion-induced collapses (to form neutron stars) rather than SNe~Ia
(e.g.\ Nomoto \& Kondo 1991; Gutierrez et~al.\ 1996). Accretion induced
collapses do not eject enough nickel to match the light curves of normal
SNe~Ia, although they may be able to explain very subluminous events like
SN~1991bg (e.g.\ Fryer et~al.\ 1999). I should note
that the existing calculations have been performed for WDs of
O--Ne--Mg composition, while some recent calculations of the evolution of a
10~M$_{\odot}$ star produce degenerate cores which are almost devoid of
magnesium (Ritossa, Garcia-Berro \& Iben 1996). Nevertheless, because of
the above two points {\it it is unlikely that O--Ne WDs produce the bulk
of SNe~Ia\/}.
\item[(iii)] {\it C--O WDs\/}: Carbon--Oxygen WDs are formed in binaries
from main sequence stars of up to $\sim10$~M$_{\odot}$. They are
therefore both relatively numerous, and they provide a significant ``phase
space volume'' (masses in the range 0.8--1.2~M$_{\odot}$; accretion
rates in the range $10^{-8}$--$10^{-6}$~M$_{\odot}$/yr) in which they
are expected to produce SNe~Ia (upon reaching the Chandrasekhar limit;
e.g.\ Nomoto \& Kondo 1991). Consequently, {\it the accreting WDs that
produce most of the SNe~Ia are very probably of C--O composition\/}!
\end{enumerate}

\subsection{At what mass does the WD explode and where and in what fuel
does the ignition take place?}

While there is virtually unanimous agreement about everything I said up
to now, namely, that: {\it SNe~Ia are thermonuclear disruptions of
accreting C--O WDs\/}, the next step in the refinement to the model is
more controversial. Two major classes of models have been considered,
and they suggest entirely different answers to the questions posed by
the title of this subsection. In one class, the WD explodes upon
reaching the {\it Chandresekhar mass\/}, as {\it carbon\/} ignites at
its {\it center\/}. In the second, the WD explodes at a {\it
sub-Chandresekhar mass\/}, as {\it helium\/} ignites {\it off-center\/}.
I will now review briefly each of these classes and point out their
strengths and weaknesses.

\subsubsection{Chandrasekhar mass carbon ignitors}

In this model, considered `standard,' the WD accretes until it
approaches the Chandrasekhar mass. Carbon ignition (triggered by
compressional heating) occurs at or very
near the center and the  burning front propagates outwards. Three types of
flame propagation models have been considered in the past three decades:
(i)~detonation (e.g.\ Arnett 1969; Hansen \& Wheeler 1969),
(ii)~deflagration (e.g.\ Nomoto, Sugimoto \& Neo 1976) and iii)~delayed
detonation, in which the flame starts as a deflagration which transitions
into a detonation at some transition density (e.g.\ Khokhlov 1991; Woosley
\& Weaver 1994). Models of the latter two types ((iii) in particular) have
generally been quite successful in explaining the observations (see e.g.\
H\"oflich \& Dominguez, these proceedings). The main
{\it strengths\/} of this model (central carbon ignition at the
Chandrasekhar mass) are (see e.g.\ H\"oflich \& Khokhlov 1996; Nugent
et~al.\ 1997; H\"oflich \& Dominguez these proceedings, for detailed
modeling):
\begin{enumerate}
\item[(1)] Some {\it $10^{51}$~ergs of kinetic energy\/} are deposited
into the ejecta by nuclear energy.
\item[(2)] $^{56}$Ni decay powers the {\it lightcurve\/}.
\item[(3)] The density and composition as a function of the ejection of
velocity (X$_{\rm i}$(V$_{\rm ej}$)) are consistent with the observed {\it
spectra\/}.
\item[(4)] The fact that the explosion occurs at the Chandrasekhar mass may
explain the broad-brush {\it homogeneity\/}.
\item[(5)] Spectra (e.g.\ of SNe~1994D, 1992A) can be fitted in great
detail by theoretical models (e.g.\ Nugent et~al.\ 1997).
\end{enumerate}

The main {\it weaknesses\/} of the Chandrasekhar mass models are:
\begin{enumerate}
\item[(1)] It has proven more difficult than originally thought for WDs
to accrete {\it up to the Chandrasekhar mass\/} in sufficient numbers
to account for the SNe~Ia rate. The difficulty is associated with mass
loss episodes in nova explosions, in helium shell flashes and in massive
winds or
common envelope phases. I will return to some of these problems when I
discuss specific progenitor models.
\item[(2)] For initial WD masses larger than $\sim1.2$~M$_{\odot}$, (which
can more easily, in principle, reach the Chandrasekhar mass) {\it
accretion-induced collapse\/} is a more likely outcome than a SN~Ia
(e.g.\ Nomoto \& Kondo 1991).
\item[(3)] {\it The late-time spectrum\/} ($\sim300$ days), and in
particular the Fe~III feature at \mbox{$\sim4700$~\AA}\ does not agree well
with Chandrasekhar mass models (Liu, Jeffrey \& Schultz 1998).
\item[(4)] The `standard' model has some difficulty in reproducing the
observed (e.g.\ Riess et~al.\ 1999a) $\apgt20$~days rise times.
\end{enumerate}

My overall assessment of Chandrasekhar mass models is that the strengths
significantly outweigh the weaknesses. The calculations of late-time,
nebular spectra involve many uncertainties, and hence I do not regard
weakness~(3) above as fatal (although clearly more work will be required
to explain it away). Both weaknesses~(1) and~(2) can be overcome if it
can be demonstrated that SNe~Ia statistics can be reproduced within the
uncertainties that still plague the theoretical population synthesis
models. As I will show in \S5, this appears indeed to be the case. Weakness
(4) can be overcome (in principle at least) by lower values of the C/O
ratio, or by the presence of a 0.2--0.4~M$_{\odot}$ envelope (see e.g.\
H\"oflich, Wheeler \& Thielemann 1998). This suggests to me that this is not
a fundamental difficulty for the model.

\subsubsection{Sub-Chandrasekhar mass helium ignitors}

In these models a C--O WD accumulates a helium layer of
$\sim0.15$~M$_{\odot}$ while the total mass is sub-Chandrasekhar. The
helium ignites off-center (at the bottom of the layer), resulting in an
event known as ``Indirect Double Detonation'' (IDD) or ``Edge Lit
Detonation'' (ELD). Basically, one detonation propagates outward
(through the helium), while an inward propagating pressure wave
compresses the C--O core which ignites off-center, followed by an
outward detonation (e.g.\ Livne 1990; Livne \& Glasner 1991; Woosley \&
Weaver 1994; Livne \& Arnett 1995; H\"oflich \& Khokhlov 1996; and
Ruiz-Lapuente, talk presented at the Chicago meeting on Type~Ia
Supernovae: Theory and Cosmology, October 1998).
\newpage

The main {\it strengths\/} of ELD (sub-Chandrasekhar) models are:
\begin{enumerate}
\item[(1)] It is easier to achieve the required {\it statistics\/},
since less mass needs to be accreted, and the WD does not need to be
extremely massive (e.g.\ Ruiz-Lapuente, Canal \& Burkert 1997;
Di~Stefano et~al.\ 1997; Yungelson \& Livio 1998).
\item[(2)] The {\it late-time spectrum\/} (in particular the Fe~III
feature at $\sim4700$~\AA) agrees better with ELD models.
\item[(3)] SNe~Ia {\it light curves\/} can be reproduced adequately by
ELD models (although the light curves rise somewhat faster than
observed, due to $^{56}$Ni heating; H\"oflich et~al.\ 1997).
\end{enumerate}

The main {\it weaknesses\/} of ELD models are:
\begin{enumerate}
\item[(1)] The {\it spectra\/} that are produced by ELD models generally
do not agree with observations (e.g.\ of SN~1994D; Nugent et~al.\ 1997).
In particular, the spectra are very blue (due to heating by radioactive Ni),
and are
dominated by Ni lines, while not showing a strong Si line.
The agreement is somewhat better for the subluminous SNe~Ia (e.g.\
SN~1991bg; Nugent et~al.\ 1997; Ruiz-Lapuente, talk presented at the
Chicago meeting on Supernovae, October 1998), but even
there it is not very good.
\item[(2)] The {\it highest velocity ejecta have the wrong composition\/}
($^{56}$Ni and He moving at 11,000 to 14,000~km~s$^{-1}$, not intermediate
mass elements; also no high velocity~C; e.g.\ Livne \&
Arnett 1995). This is due to the fact that in these models, essentially
by construction, the intermediate mass elements are sandwiched by Ni and
He/Ni rich layers, at the inner and outer sides, respectively.
\item[(3)] Since ELD models allow for a range of WD masses, and since
more massive WDs produce brighter SNe, one might expect this model to
produce a more gradual decline on the bright side of the {\it luminosity
function\/}. While this is in contradiction to the observed sharp decline
obtained for some of the earlier samples, it may not be in contradiction
with the more recent observations of a relatively flat luminosity function
(see \S2 characteristic~(3)).
\end{enumerate}

My overall assessment of the sub-Chandrasekhar mass model is that the
weaknesses (and in particular weaknesses~(1) and~(2) which appear almost
inevitable) greatly outweigh the strengths in terms of this being a
model for the bulk of SNe~Ia. It is still possible that ELDs may
correctly represent some subluminous SNe~Ia (e.g.\ Ruiz-Lapuente, Canal, \&
Burkert 1997; Pinto, private communication).

\subsection{The favored model}

On the basis of the above discussion the basic model can now be further
refined, and I tentatively conclude that: {\it Most SNe~Ia represent
thermonuclear disruptions of mass accreting C--O white dwarfs, when
these white dwarfs reach the Chandrasekhar limit and ignite carbon at
their centers\/}!

\section{The two possible scenarios}

The next step, in which we search for the progenitor systems of SNe~Ia is
even more controversial. Two possible scenarios have been proposed: (i)~The
{\it double-degenerate\/} scenario, in which two CO WDs in a binary system
are brought together by the emission of gravitational radiation and
coalesce (Webbink 1984; Iben \& Tutukov 1984). (ii)~The {\it
single-degenerate\/} scenario, in which a CO WD accretes hydrogen-rich
or helium-rich material from a non-degenerate companion (Whelan \& Iben
1973; Nomoto 1982).
\newpage

In the first scenario the progenitor systems are necessarily {\it binary
WD systems\/} in which the total mass exceeds the Chandrasekhar mass,
and which have binary periods shorter than about thirteen hours (to
allow merger within a Hubble time).

In the second scenario the progenitors could be systems like: (i)~{\it
Recurrent novae\/} (both of the type in which the WD accretes hydrogen
from a giant like T~CrB, RS~Oph, and of the type in which the WD accretes
helium rich material from a subgiant like U~Sco, V394~CrA, and Nova
LMC~1990\#2), (ii)~{\it Symbiotic Systems\/} (in which the WD accretes
hydrogen-rich material from a low mass red giant or a Mira variable), or
(iii)~persistent {\it Supersoft X-ray
Sources\/} (in which the WD accretes at a high rate
$\apgt10^{-7}$~M$_{\odot}$/yr from a subgiant companion).

I will now examine the strengths and weaknesses of each one of these
scenarios.

\subsection{The double-degenerate scenario}

There is no question that close binary white dwarf systems in which the
total mass
exceeds the Chandrasekhar mass are an expected outcome of binary star
evolution
(e.g.\ Iben \& Tutukov 1984; Iben \& Livio 1993). Once the lighter WD (which
has a larger radius) fills its Roche lobe, it is entirely dissipated within
a few orbital
periods, to form a massive disk around the primary (e.g.\ Rasio \& Shapiro
1994;
Benz, Thielemann \& Hills 1989). The subsequent evolution of the system
depends
largely on the accretion rate through this disk (e.g.\ Mochkovitch \& Livio
1990;
see discussion below).

The main {\it strengths\/} of this scenario are the following:
\begin{enumerate}
\item[(1)] The {\it absence of hydrogen\/} in the spectrum is naturally
explained in a model which involves the merger of two C--O WDs. In fact,
if hydrogen is ever detected in the spectrum of a SN~Ia, this would deal
a fatal blow to this model. Tentative evidence for circumstellar
H$\alpha$ absorption is SN~1990M was presented by Polcaro and Viotti
(1991). However, Della Valle, Benetti \& Panagia (1996) demonstrated
convincingly that the absorption was caused by the parent galaxy, rather
than by the SN environment.
\item[(2)] In spite of some impressions to the contrary, {\it many double
WD systems do exist\/}. In a sample of 153 field WDs and subdwarf
B~stars, Saffer, Livio \& Yungelson (1998) found 18 new double-degenerate
candidates. Maxted \& Marsh (1999) showed (from a radial velocity survey of
46~WDs) that there is a 95\% probability that the fraction of double
degenerates
among DA WDs lies in the range 0.017--0.19. There are currently eight known
systems with orbital periods of less than half a day (and the subdwarf
B~stars
PG~1432+159 and PG~2345+318, with orbital periods of 5.4~hr and 5.8~hr
respectively may also have WD companions; Moran et~al.\ 1999). While only
one of all of these systems (KPD~0422+5421; Koen, Orosz \& Wade (1998)) has
a
total mass which within the errors could be higher than the Chandrasekhar
mass, the sample of confirmed short-period double-degenerates is still
smaller than the number predicted to contain a massive system.
\item[(3)] Population synthesis calculations predict the {\it right
statistics\/} for mergers, about $10^{-3}$~yr$^{-1}$ events for
populations that are $\sim10^8$~yr old and $10^{-4}$~yr$^{-1}$ for
populations that are $\sim10^{10}$~yr old.
\item[(4)] Since double WD systems were found to exist, {\it mergers\/}
with some ``interesting'' consequences (either a SN~Ia or an
accretion-induced collapse) appear inevitable.
\item[(5)] The explosion or collapse is expected to occur at (or near) the
Chandrasekhar mass, which as I noted in \S4.3, I regard as a property
of the favored model.
\end{enumerate}

The main {\it weaknesses\/} of the double-degenerate scenario are the
following:
\begin{enumerate}
\item[(1)] There are strong indications that WD mergers may lead to
off-center carbon ignition, accompanied by the conversion of the C--O WD
to an O--Ne--Mg composition, followed by an accretion-induced
collapse rather than a SN~Ia (e.g.\ Mochkovitch \& Livio 1990; Saio \&
Nomoto 1985, 1998; Woosley \& Weaver 1986).
\item[(2)] Galactic chemical evolution results, and in particular the
behavior of the [O/Fe] ratio as a function of metallicity ([Fe/H]) have
been claimed to be inconsistent with WD mergers as the mechanism for
SNe~Ia (Kobayashi et~al.\ 1998).
\item[(3)] While the unusually high luminosity of SN~1991T and some of its
other features have been tentatively attributed to a super-Chandrasekhar
product of the merger of two WDs (Fisher et~al.\ 1999), there is little
evidence for example for the presence of unburned carbon (as might be
expected from the disk formed in the merger process) in most SNe~Ia.
\end{enumerate}

Since we are now getting to the final stages in the identification of
the progenitors, it is important to assess critically the severity of
the above weaknesses. I will therefore discuss now each one of them in
some detail.

\subsubsection{Constraints from Galactic chemical evolution}

Supernovae Type~II  (SNe~II) are explosions resulting from the core
collapse of massive ($\apgt8$~M$_{\odot}$) stars. These supernovae
produce relatively more oxygen and magnesium than iron ([O/Fe]~$>0$). On
the other hand, SNe~Ia produce mostly iron and little oxygen. Generally,
the impression is that metal poor stars ([Fe/H]~$\leq-1$)
have a nearly flat relation of [O/Fe] vs.\ [Fe/H], with a value of
[O/Fe]~$\sim0.45$ (e.g.\ Nissen et~al.\ 1994), while disk stars
([Fe/H]$\apgt-1$) show a linearly decreasing [O/Fe] with increasing
metallicity (e.g.\ Edvardsson et~al.\ 1993; McWilliam 1997). The
``observed''
(but see below) break (from flat to linearly decreasing) near
[Fe/H]~$\sim-1$ is traditionally explained by the fact that the early heavy
element production was done exclusively by SNe~II, with the break occurring
when the larger Fe production by SNe~Ia kicks in (e.g.\ Matteucci \& Greggio
1986).

Recently, Kobayashi et~al.\ (1998) performed chemical evolution
calculations for both the double-degenerate scenario and for the
single-degenerate scenario. For the latter they used two types of
progenitor systems: one with a red giant companion and an orbital period
of tens to hundreds of days, and the other with a near main sequence
companion and a period of a few tenths of a day to a few days.

They obtained for the double-degenerate scenario (for which they took a
time delay to the explosion of $\sim0.1$--0.3~Gyr) a break at
[Fe/H]~$\sim-2$. For the single-degenerate scenario (with a delay caused by
the main sequence lifetime of $\apgt1$~Gyr; including metallicity effects),
they obtained a break at [Fe/H]~$\sim-1$.  Kobayoshi et~al.\ (1998) thus
concluded that the Galactic chemical evolution that results from the
double-degenerate scenario is inconsistent with observations.

Personally, I am not too convinced by this apparent discrepancy, since
Galactic chemical evolution calculations and observations are
notoriously uncertain. For example, a recent determination of [Ba/Fe] as a
function of [Fe/H] shows a break near [Fe/H]~$\sim-2$, which would be
consistent with the double-degenerate scenario prediction (Burris et~al.\
1999). In addition, recent Keck observations of oxygen in unevolved
metal-poor stars appear to show no break in the [O/Fe] vs.\ [Fe/H] relation.
Rather, oxygen is enhanced relative to iron over three orders of magnitude
in [Fe/H] in a linear relation (Boesgaard et~al.\ 1999; see also Israelian,
Garcia Lopez \& Rebolo 1998). While some reservations about these findings
have been raised, in particular, a re-analysis of two of the stars of
Israelian et~al.\ shows [O/Fe] ratios which are discrepant with the results
of Israelian et~al.\ and of Boesgaard et~al.\ (Fulbright \& Kraft 1999),
this in fact demonstrates the uncertainties involved in such determinations
(see also Stephens 2000).

\subsubsection{Merger only applicable to relatively rare events?}

As I noted above, it has been shown that if SN~1991T is at the same distance
as SNe~1981B and 1960F, then its luminosity is too high to be explained in
terms of a Chandrasekhar mass ejection (Fisher et~al.\ 1999). Thus, it has
been suggested that this SN resulted from the explosion  of a
super-Chandrasekhar object, indicating perhaps that WD mergers may be
responsible for at least some SNe~Ia.

However, events like SN~1991T, which seem to be associated with regions of
active star formation, represent at most $\sim20$\% of the SNe~Ia (Li
et~al.\ 2000), and therefore, even if they are the results of mergers this
still does not mean that WD mergers are the main class of progenitors of
SNe~Ia. In addition, it is still far from clear whether mergers can lead to
explosions at all (see \S5.1.3 below). Incidentally, data for a cepheid
distance to NGC~4527 (which will help determine the true intrinsic
luminosity of SN~1991T) have been obtained with HST, and the analysis is in
progress (Saha et~al.\ 2000).

\subsubsection{SN~Ia or accretion induced collapse?}

Potentially the most serious (and possibly even fatal) weakness of the
double-degenerate scenario comes from the fact that some estimates and
calculations indicate that the coalescence of two C--O WDs may lead to
an accretion-induced collapse rather than to a SN explosion (e.g.\
Mochkovitch \& Livio 1990; Saio \& Nomoto 1985, 1998; Kawai, Saio \&
Nomoto 1987; Timmes, Woosley \& Taam 1994; Mochkovitch, Guerrero \&
Segretain 1997).

The point is the following: once the lighter WD fills its Roche lobe, it
is dissipated within a few orbital periods (Benz et~al.\ 1990; Rasio \&
Shapiro 1995; Guerrero 1994), and it forms a hot thick disk configuration
around the more massive white dwarf. This disk is mainly rotationally
supported and hence central carbon ignition does not take place
immediately, but rather the subsequent evolution depends largely on the
rate of angular momentum transport and removal, since they determine the
accretion rate onto the primary WD. As long as the accretion rate is
higher than about $\dot{M}\apgt2.7\times10^{-6}$~M$_{\odot}$ yr$^{-1}$,
{\it carbon is ignited off-center\/} (at the core-disk boundary; this
may happen during the merger itself; e.g.\ Segretain  1994). Under such
conditions, the flame was found (in spherically symmetric calculations)
to propagate all the way to the center within a few thousand years, thus
burning the C--O into an O--Ne--Mg mixture with {\it no explosion\/}
(i.e.\ before carbon is centrally ignited; e.g.\ Saio \& Nomoto 1998).
Such configurations are expected to collapse (following electron
captures on $^{24}$Mg) to form neutron stars (Nomoto \& Kondo 1991; Canal
1997). The main questions are therefore:
\begin{enumerate}
\item[(i)] What accretion rates can be expected from the initial
WD-thick disk configuration?
\item[(ii)] May some aspects of the flame propagation be different given
the fact that the real problem is three-dimensional while most of the
existing calculations were performed using a spherically symmetric code?
In particular, {\it could the carbon burning be quenched before the
transformation to O--Ne--Mg composition occurs\/}?
\item[(iii)] Could the WDs ignite even prior to the merger due to tidal
heating, and
what would be the outcome of such pre-merger ignition?
\end{enumerate}

The answers to all of these questions involve uncertainties, however
some possibilities appear more likely than others.  First, it appears {\it
very difficult to avoid high accretion rates\/}. If the MHD turbulence
that is expected to develop in accretion disks (e.g.\ Balbus \& Hawley
1998) is operative, with a corresponding viscosity parameter of
$\alpha\sim0.01$ (where the viscosity is given by $\nu\sim\alpha c_{\rm
s}H$, with $H$ being a vertical scaleheight in the disk and $c_{\rm s}$
the speed of sound; e.g.\ Balbus, Hawley \& Stone 1996), then angular
momentum can be removed in a matter of days! In such a case, even if the
accretion rate is Eddington limited (at $\sim10^{-5}$~M$_{\odot}$/yr),
off-center carbon ignition should still occur, with an eventual collapse
rather than an explosion. Deviations from spherical symmetry can only
hurt, since they may allow accretion to proceed at a super-Eddington rate.
It is difficult to see why the dynamo-generated viscosity would be
suppressed for the kind of shear and temperatures expected in the disk.

Concerning the burning itself, recent attempts at multi-dimensional
calculations of the flame propagation and a more detailed analysis of
some of the processes involved (Garcia-Senz, Bravo \& Serichol 1998;
Bravo \& Garcia-Senz 1999) indicate that if anything, accretion induced
collapses are an even more likely outcome than previously thought. This
is due to the effects of electron captures in Nuclear Statistical
Equilibrium which tend to stabilize the thermonuclear flame, and to
Coulomb corrections to the equation of state. The latter has the effect
of reducing the flame velocities and the electronic and ionic pressures,
all of which result in a reduction in the critical density which
separates explosions from collapses.

As two WDs approach merger, their interiors can be spun up by tidal forces.
{\it If\/} these tides can bring (at least one of) the WDs into quasi
synchronism between the spin period and the orbital period, high dissipation
rates and heating will ensue (Rieutord \& Bonazolla 1987). The obtained
luminosities due to this tidal heating can reach values as high as
$\apgt10^{37}$~erg~s$^{-1}$ (Rieutord \& Bonazzola 1987; Iben, Tutukov \&
Federova 1998). If such heating indeed occurs, it could have (in principle
at least) two important effects:
(i)~it would increase the probability of detection of pre-merger WDs, due to
the increased luminosity and the expected periodic variability (due to
mutual occultations), (ii)~heating could lead to carbon ignition prior to or
during the merger.

>From the point of view of the present discussion it is important to assess
whether the latter possibility makes the merging WDs more viable progenitor
systems. This does not appear to be the case for the following reasons:
\begin{enumerate}
\item[(i)] It is not obvious that a WD can be brought to synchronous
rotation. The normal viscosity of WD matter is very low (e.g.\ Durisen
1973), which can make the viscous timescale even longer than the system's
lifetime. This problem however may be overcome if turbulence develops due to
the strong shear.
\item[(ii)] It is not clear if carbon will be ignited even if tidal heating
occurs. In fact, in the calculations of Iben et~al.\ (1998) carbon failed to
be ignited (although only by a relatively narrow margin).
\item[(iii)] Even if carbon is ignited, it is very likely that the ignition
will occur off-center, making an accretion induced collapse a more likely
outcome than a SN~Ia (as explained above).
\end{enumerate}

Finally, on the observational side there are also two points which argue
at some level against WD mergers as the main SNe~Ia progenitors.
\begin{enumerate}
\item[(i)] Even if MHD viscosity could somehow be suppressed in the disk,
and the
disk surrounding the primary WD could cool down, so that angular
momentum would be transported only via the viscosity of (partially)
degenerate electrons, this would result
in an accretion timescale of $\sim10^9$~yrs (Mochkovitch \& Livio 1990;
Mochkovitch et~al.\ 1997). The system prior to the explosion would have
an absolute magnitude of $M_{\rm V}\aplt10$ (with much of the emission
occurring in the UV). There is no evidence for the existence of some
$\sim10^7$ such objects in the Galaxy.
\item[(ii)] The existence of planets around the pulsars PSR~1257+12 and
PSR~1620--26 (Wolszczan 1997; Backer 1993; Thorsett, Arzoumanian \&
Taylor 1993) could be taken to mean (this is a model dependent
statement) that mergers tend to produce accretion induced collapses rather
than SNe~Ia. In one of the leading models for the formation of such
planets (Podsiadlowski; Pringle \& Rees 1991; Livio, Pringle \& Saffer
1992), the planets form in the following sequence of events. The lighter
WD is dissipated (upon Roche lobe overflow) to form a disk around the
primary. As material from this disk is accreted, matter at the outer
edge of the disk has to absorb the angular momentum, thereby expanding
the disk to a large radius. The planets form from this disk in the
same way that they did in the solar system, while the central object
collapses to form a neutron star.
\end{enumerate}

\subsubsection{Overall assessment of the double-degenerate scenario}

It has now been observationally demonstrated that many double-degenerate
systems exist. The general agreement between the distribution of the
observed properties (e.g.\ orbital periods, masses) and those predicted
by population synthesis calculations (Saffer, Livio \& Yungelson 1998),
suggests that
the fact that no clear candidate (short period) system with a total mass
exceeding the Chandrasekhar mass has been found yet, may merely reflect
the insufficient size of the observational sample. Thus, there is very
little doubt in my mind that statistics is not a serious problem. The
most disturbing uncertainty is related to the outcome of the merger
process itself. The discussion in \S5.1.3 suggests that {\it collapse to a
neutron star is more likely than a SN~Ia\/} (see also Mochovitch et~al.\
1997).

\subsection{The single-degenerate scenario}

The main {\it strengths\/} of the single degenerate scenario are:
\begin{enumerate}
\item[(1)] A class of objects in which hydrogen is being transferred at
such high rates that it {\it burns steadily\/} on the surface of the WD
has been identified---the Supersoft X-ray Sources (e.g.\ Greiner, Hasinger
\&
Kahabka 1991; van den Heuvel et~al.\ 1992; Southwell et~al.\ 1996; Kahabka
and van den Heuvel 1997). If the accreted matter can indeed be retained,
this provides a natural path to an increase in the WD mass towards the
Chandrasekhar mass (e.g.\ Di~Stefano \& Rappaport 1994; Livio 1995, 1996a;
Yungelson et~al.\ 1996).
\item[(2)] Other candidate progenitor systems are known to exist, like
symbiotic systems (e.g.\ Munari \& Renzini 1992; Kenyon et~al.\ 1993;
Hachisu, Kato \& Nomoto 1999) and recurrent novae (Hachisu et~al. 1999a).
\item[(3)] There have been claims that the single degenerate scenario
fits better the results of Galactic chemical evolution (e.g.\ Kobayashi
et~al.\ 1998). However, as I have shown in \S5.1.1, recent observations
cast doubt on this assertion. Similarly, nucleosynthesis results show
that in order to avoid unacceptably large ratios of $^{54}$Cr/$^{56}$Fe
and $^{50}$Ti/$^{56}$Fe, the central density of the WD at the moment of
thermonuclear runaway must be lower than $\sim2\times10^9$~g cm$^{-3}$
(Nomoto et~al.\ 1997). Such low densities are realized for high
accretion rates ($\!\apgt10^{-7}$~M$_{\odot}$ yr$^{-1}$), which are
typical for the Supersoft X-ray Sources. Nucleosynthesis results suffer
too, however, from considerable uncertainties (e.g.\ Nagataki, Hashimoto
\& Sato 1998).
\end{enumerate}

The main {\it weaknesses\/} of the single degenerate scenario are:
\begin{enumerate}
\item[(1)] The upper limits on {\it radio detection\/} of hydrogen at 2~and
6~cm
in SN~1986G, taken approximately one week before optical maximum (Eck
et~al.\ 1995), rule out a symbiotic system progenitor for this system
with a wind mass loss rate of $10^{-7}\aplt \dot{M}_{\rm W}
\aplt10^{-6}$~M$_{\odot}$ yr$^{-1}$ (Boffi \& Branch 1995). This in
itself is not fatal, since SN~1986G is somewhat peculiar (e.g.\ Branch
and van den Bergh 1993), and the upper limit on the mass loss rate is at
the high end of observed symbiotic winds. An even less stringent upper
limit from x-ray and H$\alpha$ observations exists for SN~1994D (Cumming
et~al.\ 1996).
\item[(2)] There exists some uncertainty whether WDs can even reach the
Chandrasekhar mass {\it at all\/} by the accretion of hydrogen (e.g.\
Cassisi, Iben \& Tornambe 1998). Furthermore, even if they can, the
question of whether they can produce the required\pagebreak\ SNe~Ia
statistics is
highly controversial (e.g.\ Yungelson et~al.\ 1995, 1996; Yungelson \&
Livio 1998, 1999; Hachisu, Kato \& Nomoto 1999; Hachisu et~al.\ 1999a).
\end{enumerate}

I will now examine these weaknesses in some detail.

\subsubsection{Observational detection of hydrogen}

Ultimately, the presence or total absence of hydrogen in SNe~Ia will
distinguish unambiguously between single-degenerate and
double-degenerate models. To date, hydrogen has not been convincingly
detected
in {\it any\/} SN~Ia. It is interesting to note that narrow
$\lambda6300$, $\lambda6363$~[OI] lines were observed only in one SN~Ia
(SN~1937C; Minkowski 1939), but even in that case there was no hint of a
narrow H$\alpha$ line. Hachisu, Kato \& Nomoto (1999) estimate in one of
their models (which involves stripping of material from the red giant;
see below) a density measure of $\dot{M}/v_{10}\sim10^{-8}$~M$_{\odot}$
yr$^{-1}$ (where $v_{10}$ is the wind velocity in units of
10~km~s$^{-1}$), while the most stringent radio upper limit existing
currently (for SN~1986G) is $\dot{M}/v_{10}\sim10^{-7}$~M$_{\odot}$
yr$^{-1}$ (Eck et~al.\ 1995; for SN~1994D Cumming et~al.\ (1996) find from
H$\alpha$ an upper limit of $\dot{M}\sim1.5\times
10^{-5}$~M$_{\odot}$ yr$^{-1}$ for a wind speed of 10~km~s$^{-1}$; for
SN~1992A Schlegel \& Petre (1993) find from X-ray observations an upper
limit of $\dot{M}/v_{10}=(2-3)\times10^{-6}$~M$_{\odot}$ yr$^{-1}$). Thus,
while it is impossible at present to rule out single-degenerate models on
the basis of the apparent absence of hydrogen, the hope is that near
future observations will be able to determine definitively whether this
absence is real or if it merely represents the limitations of existing
observations (an improvement by two orders of magnitude in the sensitivity
will give a
definitive answer). I should note that a narrow emission feature possibly
corresponding to H$\alpha$ was detected in SN~1981b (in NGC~4536), however
no trace of the emission was seen 5~days later (Branch et~al.\ 1983).

\subsubsection{Statistics}

Growing the WD to the Chandrasekhar mass is not easy. At accretion rates
below $\sim10^{-8}$~M$_{\odot}$/yr WDs undergo repeated nova outbursts
(e.g.\ Prialnik \& Kovetz 1995), in which the WDs lose more mass than
they accrete between outbursts (e.g.\ Livio \& Truran 1992). For
accretion rates in the range $10^{-8}$--a few
$\times10^{-7}$~M$_{\odot}$/yr,
while helium can accumulate, the WDs experience mass loss due to helium
shell flashes and due to the common envelope phase which results from
the engulfing of the secondary star in the expanding envelope (with mass
loss
occurring due to drag energy deposition). At accretion rates above a few
$\times10^{-7}$~M$_{\odot}$/yr, the WDs expand to red giant configurations
and lose mass due to drag in the common envelope and due to winds (e.g.\
Cassisi et~al.\ 1998). The net result of these constraints has been that
population
synthesis calculations which follow the evolution of all the binary systems
in the Galaxy, tended until recently to conclude that single degenerate
channels manage to bring WDs to the Chandrasekhar mass only at about
10\% of the inferred SNe~Ia frequency of $4\times10^{-3}$~yr$^{-1}$ (e.g.\
Yungelson et~al.\ 1995, 1996; Yungelson \& Livio 1998; Di~Stefano et~al.\
1997; although see Li \& van den Heuvel 1997).

Very recently, a few serious attempts have been made to investigate whether
the statistics could be improved by increasing the ``phase space'' for
single degenerate scenarios, given the fact that population synthesis
calculations involve many assumptions. These attempts resulted in the
identification of three
directions in which the phase space could (potentially) be increased.
\begin{enumerate}
\item[(i)] The accumulation efficiency of helium has been recalculated
using OPAL opacities (Kato \& Hachisu 1999). These authors concluded
that helium can accumulate much more efficiently than found by Cassisi
et~al.\ (1998), mainly because the latter authors used relatively low
WD masses (0.516~M$_{\odot}$ and 0.8~M$_{\odot}$) and old opacities in
their calculations.
\item[(ii)] Hachisu et~al.\ (1999a,b) claimed to have identified two
evolutionary channels for single-degenerate systems previously overlooked
in population synthesis calculations. In the first of these channels, the
C--O WD is
formed from a red giant with a helium core of 0.8--2.0~M$_{\odot}$
(rather than from an asymptotic giant branch star with a C--O core). The
immediate progenitors in this case are expected to be either helium-rich
Supersoft X-ray Sources or recurrent novae of the U~Sco subclass (where the
accreted material appears to be helium rich).
\item[]In the second channel, Hachisu et~al.\ (1999b) considered very wide
(initial
separations as large as $\sim40,000$~R$_{\odot}$) symbiotic systems, in
which
the components are brought together by the inclusion of new physical effects
(see
(iii)--(3) below).
\item[(iii)] It has been suggested that the inclusion of a few
additional physical effects, can increase substantially the phase space
of the symbiotic channel (Hashisu, Kato \& Nomoto 1996, 1999b). These new
effects included:
\end{enumerate}
\begin{enumerate}
\item[(1)] The WD loses much of the transferred mass in a massive wind.
This has the effect that the mass transfer process is stabilized for a
wider range of mass ratios, up to $q_{\rm max}\equiv m_2/m_1=1.15$ instead
of $q_{\rm max}=0.79$ without the massive wind.
\item[(2)] It has been suggested that the wind from the WD strips the
outer layers of the red giant at a high rate. This increases the allowed
mass ratios (for stability) even above 1.15, essentially indefinitely.
\item[(3)] It has been suggested that at large separations (up to
$\sim40,000$~R$_{\odot}$), when the orbital velocity is of the order of the
wind
velocity, the wind from the red giant acts like a common envelope to reduce
the
separation, thus allowing much wider initial separations to result in
interaction.
\end{enumerate}

There are many uncertainties associated with all of these attempts to
increase the phase space. For example, the efficiency of mass stripping
from the giant by the wind from the WD may be much smaller than assumed
by Hachisu et~al.\ (1999b), for the following reasons. At high accretion
rates, much of the mass loss from the WD may be in the form of an
outflow or a collimated jet, perpendicular to the accretion disk rather
than in the direction of the giant. Evidence that this is the case is
provided by the jet satellite lines to He~II~4686, H$\beta$ and H$\alpha$
observed in the Supersoft X-ray Source RX~J0513.9$-$6951 (Southwell et~al.\
1996). These jet lines are very similar to those seen in the
prototypical jet source SS~433 (e.g.\ Vermeulen et~al.\ 1992).
Furthermore, even if some of the WD wind hits the surface of the giant,
it is not clear how efficient it would be in stripping mass, since the
rate of energy deposition per unit area by the wind is smaller by two
orders of magnitude that the giant's own intrinsic flux.

Similarly, the efficiency of helium accumulation is still highly
uncertain, as the differences between the results of Kato \& Hachisu
(1999) and Cassisi et~al.\ (1998) have shown.

Also, the particular form of the specific angular momentum in the wind used
by Hachisu, Kato \& Nomoto (1999b; point (3) above) may be realized only in
relatively rare cases (Yungelson \& Livio 2000).

Finally, all the new suggestions for the increase in phase space rely very
heavily
on the results of the wind solutions of Kato (1990; 1991), which involve a
treatment
of the radiation and hydronamics not nearly as sophisticated as that of more
state
of the art radiative transfer codes (e.g.\ Hauschildt et~al.\ 1995, 1996).

\subsubsection{Overall assessment of the single-degenerate scenario}

The above discussion suggests that probably not all the scenarios for
increasing the ``phase space'' of the single-degenerate channels work
(if they did, we might have had the opposite problem of too high a
frequency of SNe~Ia!).  However, these attempts serve to demonstrate
that {\it the input physics to population synthesis codes still involves
many
uncertainties\/}. My feeling is therefore that given the many potential
channels leading to SNe~Ia, statistics should not be regarded as a
serious problem.

Single-degenerate scenarios therefore appear quite promising, since
unlike the situation a decade ago, a class of objects in which the WDs
accrete hydrogen steadily (the Supersoft X-Ray Sources) has actually
been identified. The main problem with single-degenerate scenarios
remains the non-detection of hydrogen so far. While a difficult
observational problem (see \S6), {\it the establishment of the presence or
absence of hydrogen in SNe~Ia should become a first priority for SNe
observers\/}.

\subsection{What if....?}

Given the fact that there are still uncertainties involved in identifying
the
SNe~Ia progenitors, and that WD mergers and some form of off-center
helium ignitions almost certainly occur, it is instructive to pose a few
``what if'' questions. For example:  {\it What if WD mergers with a
total mass exceeding Chandrasekhar do not produce SNe~Ia, what do they
produce then\/}? The answer in this case will have to be that they
almost certainly produce either neutron stars via accretion induced
collapses, or single WDs, if the merger is accompanied by extensive mass
loss from the system. Fryer et~al.\ (1999) estimate (from nucleosynthesis
constraints) that less than 0.1\% of the total Galactic neutron star
population is produced via accretion induced collapses.

{\it What if off-center helium ignitions (ELDs) do not produce SNe~Ia\/}? In
this case, if an explosive event indeed ensues, a population of ``super
novae'' (with $\sim0.15$~M$_{\odot}$ of $^{56}$Ni and He) is yet to be
detected (maybe SN~1885A in M31 was such an event?).  {\it What if
off-center helium ignitions do produce SNe~Ia? What comes out of the systems
with $M_{\rm WD}\apgt1$~M$_{\odot}$, which should be even brighter\/}?
If indeed $\sim20$\% of SNe~Ia are SN~1991T-like (Li et~al.\ 2000), then
maybe these could be represented by such events. This is far from certain,
however, since an analysis of the properties of SN~1991T showed that these
properties would be very difficult to reproduce even with a nickel mass
approaching the Chandrasekhar mass (Fisher et~al.\ 1999). Thus, we see that
off-center helium ignitions seem to present an observational problem both if
they {\it do\/} and if they {\it do not\/} produce SNe~Ia. To me this
suggests that the physics of these events is not yet well understood (for
example, maybe off-center helium ignition fails to ignite the C--O core
after all).

\section{How can we hope to unambiguously identify the progenitors?}

There are several ways in which observations of both nearby and distant
supernovae could
solve the mystery of SNe~Ia progenitors:
\begin{enumerate}
\item[(1)] A combination of {\it early high resolution optical
spectroscopy, x-ray observations\/} and {\it radio observations\/} of nearby
SNe~Ia can
both provide limits on $\dot{M}/v$ from the progenitors and potentially
detect the presence of circumstellar hydrogen (if it exists).

\item []For example, narrow HI in emission or absorption could be detected
either very early, or shortly after the ejecta become optically thin
($\sim100$~days). The latter is true because the SN ejecta probably
engulfs the companion at early times (e.g.\ Chugai 1986; Livne, Tuchman
\& Wheeler 1992). The interaction of the ejecta with the circumstellar
medium can be observed either in the radio (e.g.\ Boffi \& Branch 1995)
or in x-rays (e.g.\ Schlegel 1995). The collision of the ejecta (with
circumstellar matter) can also set up a forward and a reverse shock
(e.g.\ Chevalier 1984; Fransson, Lundqvist \& Chevalier 1996), and
radiation from the latter can ionize the wind and produce H$\alpha$
emission (e.g.\ Cumming et~al.\ 1996).
\item[(2)] Early observations (again of nearby SNe~Ia) of the gamma-ray
light curve (or gamma-ray
line profiles) could distinguish between carbon ignitors and
sub-Chandrasekhar helium ignitor models (see \S4.2.2) since the  latter
can be expected to result in a quicker rise of the gamma-ray light curve
due to the presence of $^{56}$Ni in the outer layers (and different
gamma-ray line profiles; because of the high velocity $^{56}$Ni).
\item[(3)] Another important aspect of the single degenerate scenario which
can be tested by observations  of both nearby and very distant supernovae is
the dependence on metallicity. The increase in the ``phase space'' of single
degenerate progenitors, which is required to make the statistics more
compatible with observations (\S5.2.2), relies heavily on the existence of
an optically thick wind from the WD. For a low metallicity of the accreted
matter ([Fe/H]$\aplt-1$), the wind from the WD is strongly suppressed (since
the wind is driven by a peak in the opacity, which is due to iron lines;
e.g.\ Kobayashi et~al.\ 1998). Consequently, it is expected that SNe~Ia
rates will be significantly lower in low iron abundance environments. Thus,
determinations of relative  rates in dwarf galaxies (and in the very
outskirts of spiral galaxies) can help determine the viability of a key
ingredient in the single degenerate scenario.
\item[] Similarly, a significant drop may be expected in the rate of SNe~Ia
at redshift $z\sim2$ (again due to the decrease in metallicity).
\item[(4)] In general, observations of very distant supernovae (at
$z\sim2$--4) with
the Next Generation Space Telescope (NGST) can help significantly in
identifying the progenitors (e.g.\
Yungelson \& Livio 1999, 2000; Nomoto et~al.\  2000). For example, the
progenitors can be identified
from the observed frequency of SNe~Ia as a function of redshift (e.g.\
Yungelson \& Livio 1998, 1999, 2000; Ruiz-Lapuente \& Canal 1998; Madau,
Della
Valle \& Panagia 1998; Nomoto et~al.\ 2000; and see \S8), since different
progenitor models
produce different redshift distributions. Personally, I think that it
would be quite pathetic to have to resort to this possibility.
Rather, one would like to be able to identify the progenitors independently,
and then
use the observations of supernovae at high~$z$ to constrain models of cosmic
star formation rates, and of
cosmic evolution of SNe rates, luminosity, and input into galaxies.
\end{enumerate}

\section{Cosmological implications: could we be fooled?}

One of the key questions that result from the uncertainties in the
theoretical models and the fact that we do not know with certainty which
systems are the progenitors of SNe~Ia is clearly: {\it
is it possible that SNe~Ia at higher redshifts are systematically dimmer
than their low-redshift counterparts\/}? In this respect it is important
to remember that a systematic decrease in the brightness by $\sim0.25$
magnitudes is sufficient to explain away the need for a cosmological
constant. This question became particularly relevant when an analysis of the
rise times of SNe~Ia (which was based on preliminary estimates for the
high-$z$ sample; by Goldhaber 1998 and Groom 1998) seemed to show that
high-redshift SNe have shorter rise times by 2.5~days than the low-$z$ SNe
(Riess et~al.\ 1999b). A more recent analysis (which used more realistic
error estimates than those used by Goldhaber 1998), however,  found a better
agreement (within 2$\sigma$) between the rise times of the low- and
high-redshift SNe~Ia (Aldering, Knop \& Nugent 2000).

In a recent work, Yungelson \& Livio (1999) calculated
the expected ratio of the rate of SNe~Ia to the rate of SNe from massive
stars
(Types~II, Ia, Ic) as a function of redshift for {\it several\/} progenitor
models.
{\it The possibility of having different classes of progenitors contributing
to the total SNe~Ia rate should definitely be considered\/}, especially in
view of the tentative finding by Li et~al.\ (2000) that there is a
relatively high rate ($\sim40$\%) of peculiar SNe~Ia among the local sample.
If confirmed, these findings suggest that homogeneity should no longer be
considered a very strong constraint on progenitor models. I should note
though that diversity among SNe~Ia does not {\it necessarily\/} imply
different progenitors, since even in the context of one progenitor model
diversity may arise for example from changes in the carbon mass fraction of
the WD, which in turn may depend on the environment (e.g.\ Nomoto et~al.\
2000). Yungelson and Livio (1999) showed (within the uncertainties of
population synthesis models) that {\it if\/} different progenitor systems
can contribute to the total SNe~Ia rate (e.g.\ double-degenerates and single
degenerates), then it is possible, in principle, that one class of
progenitors (e.g.\ double degenerates) will dominate the rates of the local
(low-$z$) sample,  while a different progenitor class (e.g.\ single
degenerate) will start to dominate at $z\aplt1$. This is a consequence of
the fact that the SNe~Ia rate from double degenerates is expected to decline
quite steeply from $z=0$ to $z\sim1$, while the rate from single degenerates
is expected to stay relatively flat in this redshift interval. However, at
least within the assumptions of their model calculations, it appears that
such a transition is not very likely, because of the following reason: If
the contribution from physically different channels (like double degenerates
and single degenerates, both at the Chandrasekhar mass) was indeed
significant, with a transition from dominance by double degenerates to
single degenerates occurring at $z\aplt1$, one would have expected to
observe this division more clearly in the local and distant samples. For
example, the local sample should be dominated by the double degenerate
progenitors (with a ratio of double to single degenerates which should be
consistent with the results of Li et~al.\ 2000). At the same time, however,
the high-$z$ ($z\aplt1$) sample should be dominated by single degenerates,
but with the contributions from the single and double degenerate channels
not being vastly different (in particular, the contribution should be equal
at the transition point). This is {\it not\/} consistent with the
observations of the high-$z$ sample, the latter appearing to be (within the
observational uncertainties) {\it very homogeneous\/} (Li et~al.\ 2000).
Consequently, I do not think it likely that the observed universal
acceleration is an artifact of the observed SNe~Ia sample being dominated by
different progenitor classes at high- and low-$z$ (this view is supported by
the measurements of the anisotropy of the microwave background).

I should note that the most surprising aspect in the results of Li et~al.\
(2000) is the fact that although very bright SNe~Ia (SN~1991T-like)
constitute ($21\pm7$)\% of the local sample, these bright objects appear to
be totally absent from the high-$z$ sample. One way in which one could (in
principle) explain this fact is the following.  Suppose that the
SN~1991T-like events are caused by mergers (since a super-Chandrasekhar mass
is possible in this case), while the ``normals'' are caused by single
degenerates. In this case the local sample would be dominated by single
degenerates ($\sim60$\% of the events to agree with Li et~al., 2000), while
double degenerates would contribute $\sim20$\% of the events (I ignore here
the weak events since they cannot be seen at high-$z$). Now, since the rate
of events from single degenerates stays quite flat till $z\sim1$, while the
rate of events from double degenerates declines quite steeply towards
$z\sim1$, the bright objects will be missing from the high-$z$ sample. Note,
however, that this potential explanation for the behavior of the diversity
has no obvious implications for the finding of accelerating expansion, since
the same class of progenitors dominates both the high-$z$ and low-$z$
samples. Nevertheless, a better understanding of the apparent diversity at
low-$z$ and apparent lack thereof at high-$z$ is definitely needed.

Other evolutionary effects are still possible, in principle (e.g.\ Drell,
Loredo \& Wasserman, 1999; Hillebrandt 2000), however, as far as I am aware,
only one that is physically meaningful, likely, and mimics accelerated
expansion, has been identified so far.

This one potential evolutionary effect that certainly deserves more work is
{\it the effect of metallicity on the density at the point of carbon
ignition\/}. Generally, it is expected that a lower metallicity will result
in a lower central density (e.g.\ Nomoto et~al.\ 1997). This is because a
lower metallicity results in a lower abundance of the Urca-active element
$^{21}$Ne, which in turn reduces the neutrino cooling and leads to an
earlier ignition. A lower central density could (in principle at least)
result in a more rapid light curve development (due to the lower WD binding
energy), and a lower inferred maximum brightness.

It is important to note that in a recent work, Riess et~al.\ (2000) have
shown that it is highly unlikely that the dimming of distant SNe~Ia is
caused by dust opacity (Galactic-type dust was rejected at the 3.4$\sigma$
confidence level, and ``gray'' dust with grain size $>0.1~\mu$m was rejected
at the 2.3 to 2.6$\sigma$ confidence level).

\section{Tentative conclusions and observational tests}

On the basis of the analysis and discussion in the present work, the
following tentative conclusions can be drawn:
\begin{enumerate}
\item[(1)] SNe~Ia are almost certainly thermonuclear disruptions of mass
accreting {\it C--O white dwarfs\/}.
\item[(2)] It is very likely that the explosion occurs {\it at the
Chandrasekhar mass\/}, as {\it carbon is ignited at (or very near) the WD
center\/}.
The flame propagates either as a deflagration, or, more likely perhaps,
starting as a deflagration which transitions into a detonation. Off-center
ignition of helium at sub-Chandrasekhar masses may still be
responsible for a subset of the SNe~Ia which are subluminous, but this is
less clear.
\item[(3)] The immediate progenitor systems are still not known with
certainty. From the discussion in \S5 (see in particular \S5.1.3 and
5.2.3) however, I conclude that presently {\it single degenerate
scenarios look more promising\/}, with hydrogen or helium rich material
being transferred from a subgiant or giant companion (systems like
Supersoft X-Ray Sources and Symbiotics). It is still possible, however, in
view of the apparent diversity in the local sample, that more than one
progenitor class contributes to the total SNe~Ia rate. In particular, a
scenario in which single degenerates contribute $\apgt60$\% of the events
and double degenerates $\sim20$\%, appears to be consistent with the
diversity of the $z\sim0$ sample and the lack thereof in the distant sample.
\item[(4)] Definitive answers concerning the nature of the progenitors
can be obtained from observations taken as early as possible in: {\it
x-rays, radio, and high resolution optical spectroscopy. The
establishment of the presence or absence of hydrogen in SNe~Ia should be
regarded as an extremely high priority goal for supernovae observers\/}.
If hydrogen will not be detected at interesting limits (corresponding to
$\dot{M}/v_{10}\sim10^{-8}$~M$_{\odot}$ yr$^{-1}$), this will point
clearly towards the double-degenerate scenario.
\item[(5)] Observations of SNe~Ia at high redshifts can help to test
particular ingredients of the models which are directly related to the
nature of the progenitors. For example, most of the models aiming at
improving the statistics of the single-degenerate scenarios rely on a
strong wind from the accreting WD. These models thus predict an
``inhibition'' of SNe~Ia in low-metallicity environments, and in
particular a significant decrease in the rate of SNe~Ia  in spirals at
$z\sim2$ (Kobayashi et~al.\ 1998; Nomoto et~al.\ 2000). Furthermore, if the
inferred cosmic star formation rate is used (e.g.\ Pettini et~al.\ 1998),
then the SN~Ia rate is expected to drop significantly at $z\sim1.6$. At
present, the detection of a very likely SN~Ia at redshift $z=1.32$
(SN~1997ff; Gilliland, Nugent \& Phillips 1999) in the Hubble Deep Field,
and two more at redshifts 1.20 and 1.23 (Perlmutter et~al., private
communication and Tonry et~al., private communication) appear to be at least
mildly inconsistent with this prediction, but more observations will be
required to give a more definitive answer.
\item[(6)] The potential ``inhibition'' of SNe~Ia due to low metallicity
should also manifest itself in an absence (or at least a significant decline
in the rate) of SNe~Ia in dwarf galaxies and in the outer regions of
spirals. The statistics necessary to test this prediction are starting to
accumulate.
\item[(7)] It is possible, in principle, that the local and high-$z$ SNe~Ia
samples are dominated by different progenitor classes, thus mimicking
accelerated expansion, however, this is neither very likely nor consistent
with the observations of diversity. {\it With more detections of SNe~Ia at
redshifts $z\apgt1$ it will probably become possible to directly confirm the
transition in the expansion of the universe from deceleration to
acceleration\/}. Such a transition would be difficult to mimic by systematic
or evolutionary effects, and it would therefore confirm the accelerated
expansion.
\end{enumerate}

\begin{acknowledgments}
This research has been supported in part by NASA Grant NAG5--6857.

I am grateful to Adam Riess for helpful discussions.

\end{acknowledgments}

\end{document}